\documentclass[aps,showpacs,showkeys]{revtex4}
\usepackage{amsmath,amsthm,amssymb,epsfig,alltt}

\begin{document}

\def\a{\alpha}
\def\b{\beta}
\def\d{{\delta}}
\def\l{\lambda}
\def\e{\epsilon}
\def\p{\partial}
\def\m{\mu}
\def\n{\nu}
\def\t{\tau}
\def\th{\theta}
\def\s{\sigma}
\def\g{\gamma}
\def\o{\omega}
\def\r{\rho}
\def\z{\zeta}
\def\D{\Delta}
\def\G{\Gamma}
\def\half{\frac{1}{2}}
\def\hatt{{\hat t}}
\def\hatx{{\hat x}}
\def\hatp{{\hat p}}
\def\hatX{{\hat X}}
\def\hatY{{\hat Y}}
\def\hatP{{\hat P}}
\def\haty{{\hat y}}
\def\whatX{{\widehat{X}}}
\def\whata{{\widehat{\alpha}}}
\def\whatb{{\widehat{\beta}}}
\def\whatV{{\widehat{V}}}
\def\hatth{{\hat \theta}}
\def\hatta{{\hat \tau}}
\def\hatrh{{\hat \rho}}
\def\hatva{{\hat \varphi}}
\def\barx{{\bar x}}
\def\bary{{\bar y}}
\def\barz{{\bar z}}
\def\baro{{\bar \omega}}
\def\barpsi{{\bar \psi}}
\def\sp{\sigma^\prime}
\def\nn{\nonumber}
\def\cb{{\cal B}}
\def\2pap{2\pi\alpha^\prime}
\def\wideA{\widehat{A}}
\def\wideF{\widehat{F}}
\def\beq{\begin{eqnarray}}
 \def\eeq{\end{eqnarray}}
 \def\4pap{4\pi\a^\prime}
 \def\op{\omega^\prime}
 \def\xp{{x^\prime}}
 \def\sp{{\s^\prime}}
 \def\ap{{\a^\prime}}
 \def\tp{{\t^\prime}}
 \def\zp{{z^\prime}}
 \def\xpp{x^{\prime\prime}}
 \def\xppp{x^{\prime\prime\prime}}
 \def\barxp{{\bar x}^\prime}
 \def\barxpp{{\bar x}^{\prime\prime}}
 \def\barxppp{{\bar x}^{\prime\prime\prime}}
 \def\zetap{{\zeta^\prime}}
 \def\barchi{{\bar \chi}}
 \def\baro{{\bar \omega}}
 \def\bpsi{{\bar \psi}}
 \def\barg{{\bar g}}
 \def\barz{{\bar z}}
 \def\bareta{{\bar \eta}}
 \def\ta{{\tilde \a}}
 \def\tb{{\tilde \b}}
 \def\tc{{\tilde c}}
 \def\tz{{\tilde z}}
 \def\tJ{{\tilde J}}
 \def\tpsi{\tilde{\psi}}
 \def\tal{{\tilde \alpha}}
 \def\tbe{{\tilde \beta}}
 \def\tga{{\tilde \gamma}}
 \def\tchi{{\tilde{\chi}}}
 \def\barth{{\bar \theta}}
 \def\bareta{{\bar \eta}}
 \def\barom{{\bar \omega}}
 \def\bole{{\boldsymbol \epsilon}}
 \def\bolth{{\boldsymbol \theta}}
 \def\bomega{{\boldsymbol \omega}}
 \def\bolmu{{\boldsymbol \mu}}
 \def\bolal{{\boldsymbol \alpha}}
 \def\bolbe{{\boldsymbol \beta}}
 \def\bolL{{\boldsymbol  L}}
 \def\bolX{{\boldsymbol X}}
 \def\boln{{\boldsymbol n}}
 \def\bols{{\boldsymbol s}}
 \def\bolS{{\boldsymbol S}}
 \def\bola{{\boldsymbol a}}
 \def\bolA{{\boldsymbol A}}
 \def\bolJ{{\boldsymbol J}}
 \def\tr{{\rm tr}}

\setcounter{page}{1}
\title[]{Four-graviton scattering and string path integral in the proper-time gauge}

\author{Taejin Lee}
\email{taejin@kangwon.ac.kr}
\affiliation{Department of Physics, Kangwon National University, Chuncheon 24341 Korea}

\date{\today }

\begin{abstract}
We evaluate the four-closed-string scattering amplitude, using the Polyakov string path integral in the proper-time gauge. By identifying the Fock space representation of the four-closed-string-vertex, we obtain a field theoretic expression of the closed string scattering amplitudes. In the zero-slope limit, the four-closed-string scattering amplitude reduces to the four-graviton-scattering amplitude of Einstein’s gravity. However, at a finite slope, the four-graviton scattering amplitude in the proper-time gauge differs not only from that of Einstein gravity, but also significantly differs from the conventional one obtained by using the vertex operator technique in string theory.
This discrepancy is mainly due to the presence of closed string tachyon poles in the four-graviton-scattering amplitude, which are missing in previous works. Because the tachyon poles in the scattering amplitude considerably alter the short distance behavior of gravitational interaction, 
they may be important in understanding problems associated with the perturbative theory of quantum gravity and the dark matter within the framework of string theory. 
\end{abstract}




\maketitle

\section{Introduction}

The development of a finite consistent quantum theory of gravity \cite{Ashtekar1974,Carlip2001} is one of the most outstanding problems in theoretical physics;
for this, many theorists have suggested the string theory. Because the spin-two massless particle, corresponding to the graviton, is 
included in the spectrum of a free closed string, closed string field theory may be a candidate for a unifying framework for the finite quantum theory of gravity. However, despite decades of efforts, development of a consistent closed string field theory has not been accomplished. 

In a recent work \cite{TLeeEPJ2018}, we evaluated the three-closed string scattering amplitude by choosing the proper-time gauge 
\cite{TLeeann1988,TLeePLB2017,TLeeJKPS2017,Lee2017d,TLee2017cov} for the Polyakov string path integral \cite{Polyakov1981} that depicts the three-closed string scattering. In the proper-time gauge, the string path integral can be written as integrals over the proper-times in a way similar to the Schwinger's proper time representation \cite{Schwinger1951} of Feynman integrals of quantum field theory. 
Hence, it is feasible to use the proper-time gauge to identify the field theoretical expressions of the string path integrals that depict multiple string scatterings.
If we evaluate the string path integral on a cylindrical surface by choosing the proper-time gauge, we may obtain the free field action of closed string \cite{TLeeann1988}. The Fock space representation of the three-closed string vertex may be obtained by recasting the 
string path integral on the string worldsheet, called the pants diagram, into the corresponding field theoretical expression \cite{TLeeEPJ2018}. The three-closed string scattering amplitude was found to be factorized entirely into those of the three-open string 
scattering amplitudes. This implies that the Kawai-Lewellen-Tye (KLT) \cite{Kawai1986} relations of the first quantized string theory may be extended to the second quantized string theory. 

In this work, we study the four-closed string scattering amplitude by extending the previous work on the three-closed string scattering amplitude. The graviton scattering amplitudes have been important topics in string theory. By comparing graviton scattering amplitudes evaluated in string theory with 
those of Einstein’s gravity, we can understand the relation between the perturbative quantum theory of
closed string and the Einstein gravity. Yoneya first showed \cite{Yoneya1973,Yoneya1974prog} that the closed string theory, referred to as the Virasoro--Shapiro model, contains the Einstein gravity as a zero-slope limit, by calculating the scattering amplitudes of two gravitons and two scalar particles. His work was further extended by Scherk and Schwarz \cite{Scherk1974PLB,Scherk1974NPB} to multi-graviton scattering amplitudes. These works in the early days of the string theory employed vertex operator technique to represent the external strings by point-like operators, called vertices. However, in a recent work \cite{TLeeGauge2018}, it was pointed out that if the 
vertex operator technique is adopted, the tachyon poles are missed in four-gauge-particle-scattering amplitude in open string theory. As a result, the scattering amplitudes obtained by the vertex operator technique significantly differ from those obtained without using any approximations of the Polyakov string path integral in the proper-time gauge. We expect that this discrepancy
may also exist in closed string theory, because the closed string scattering amplitudes 
are factorized into those of open string scattering amplitudes, satisfying KLT relations. 

The Polyakov string path integral is evaluated by constructing a closed string Green's function on the worldsheet. We map the worldsheet of four-closed-string-scattering onto the complex plane and rewrite the string path integral in the proper-time gauge in terms of the oscillatory basis to obtain the Fock space representation of the four-closed-string vertex. If a four-graviton state is chosen as the external four-closed string state, the Fock space representation of the four-closed-string vertex yields the four-graviton-scattering amplitude. The resultant four-graviton-scattering amplitude is found to contain the tachyon poles in all three channels, in contrast to the conventional one obtained by the vertex operator technique \cite{Schwarz1982}. 
In the zero-slope limit, it reduces to the four-graviton-scattering amplitude of Einstein gravity \cite{DeWitt1967} similar to the conventional one \cite{Sannan1986}. However, at a finite slope, the four-graviton scattering amplitude in the proper-time gauge differs not only from that of Einstein’s gravity but also significantly from the conventional one in string theory obtained using the vertex operator technique. 

\section{String Path Integral for Four-Closed String Scattering}

The four-closed string scattering amplitude is given by the string path integral of
Polyakov on the string worldsheet with four external closed strings \cite{GreenSW1987}
\begin{subequations}
\beq
{\cal W}_{[4]} &=& g^2 \int D[X]D[h] \exp \left(iS + i \int_{\p M} \sum_{r=1}^4  P^{(r)}\cdot X^{(r)} d\s \right), \label{W4}\\
S &=& -\frac{1}{4\pi} \int_M d\t d\s \sqrt{-h} h^{\a\b} \frac{\p X^\m}{\p \s^\a} \frac{\p X^\n}{\p \s^\b} \eta_{\m\n}, ~~~~~ \m, \n = 0, \dots , d-1  
\eeq
\end{subequations}
where 
$\s^1= \t$, $\s^2 = \s$. Here, $d = 26$ for the bosonic string and $d = 10$ for the open super-string. In terms of normal modes, the string coordinate variable, $X(\t,\s) = X_L(\t+\s) + X_R (\t-\s)$ may be expanded as 
\begin{subequations}
\beq
X_L (\t,\s) &=& x_L + \sqrt{\frac{1}{2}} \, p_L(\t+\s) + i \sqrt{\frac{1}{2}} \,\sum_{n\not=0} \frac{1}{n}
\a_n e^{-in(\t+\s)}, \\
X_R (\t,\s) &=& x_R + \sqrt{\frac{1}{2}} \, p_R(\t-\s) + i \sqrt{\frac{1}{2}} \,\sum_{n\not=0} \frac{1}{n}
\ta_n e^{-in(\t-\s)}.
\eeq 
\end{subequations}

\begin{figure}[htbp]
   \begin {center}
    \epsfxsize=0.8\hsize

	\epsfbox{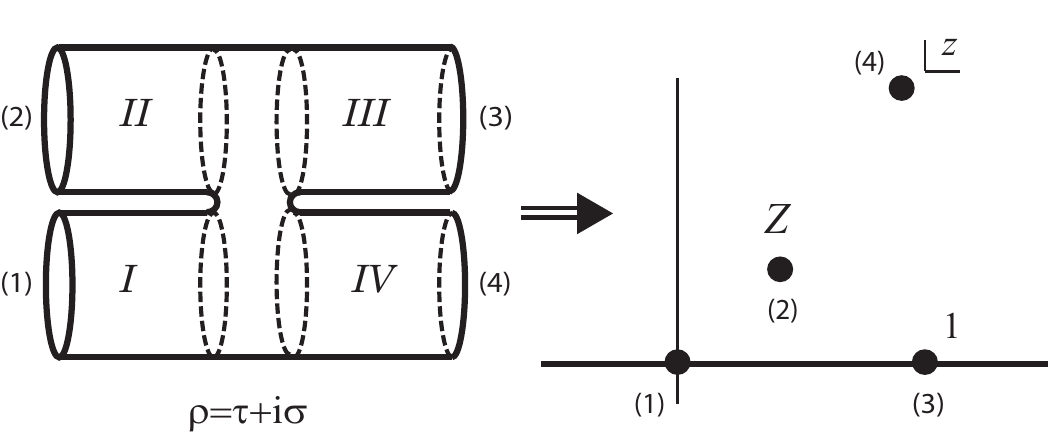}
   \end {center}
   \caption {\label{fourclosed} Local coordinate patches and the Schwarz-Christoffel mapping from the string worldsheet onto the complex plane}
\end{figure}

Fig.\ref{fourclosed} depicts the worldsheet of the four-closed string scattering on which we introduce the local coordinate patches. The two-dimensional worldsheet metric may be written in terms of the lapse and shift functions as follows
\beq
\sqrt{-h} \left(h^{\a\b}\right) = \frac{1}{N_1} \begin{pmatrix} -1 & N_2 \\ N_2 & (N_1)^2 - (N_2)^2 \end{pmatrix}.
\eeq
In the proper-time gauge, the worldsheet metric on a local patch is fixed by two constants $n_1$ and $n_2$, which correspond to the zero modes of the lapse and shift functions, respectively. The proper time $s$ on a local patch is given as $s = n_1 \Delta \t$,
where $\Delta \t$ is the interval of the local patch along $\t$ direction.  

By using the reparametrization invariance of the Polyakov string path integral, we can fix the length parameters as 
$\a_1 = \a_2 =1$ and $\a_3 = \a_4 =-1$. To evaluate the string path integral, we map the worldsheet of the four-closed string scattering onto the complex plane. If we choose the Koba--Nielsen variables as $Z_1 =0, Z_2 = Z, Z_3 =1, Z_4 =\infty$, the Schwarz--Christoffel (SC) tansformation from the worldsheet onto the complex plane is given by \beq
\rho = \ln z + \ln (z-Z) - \ln (z-1) + i\pi .
\eeq 
Two interaction points where two closed strings join to form one closed string or where one closed string splits into two 
closed strings are mapped to two points $(\t_i, \s_i)$, $i=1, 2$ on the complex plane, which are determined by 
$\frac{\p \rho}{\p z} = 0$: 
\begin{subequations}
\beq
\t_1 &=& 2 \,\text{Re}\, \ln \left(1- \sqrt{1-Z} \right), ~~\s_1 = 2 \,\text{Im}\, \ln \left(1- \sqrt{1-Z} \right) + \pi , \\
\t_2 &=& 2 \,\text{Re}\, \ln \left(1+ \sqrt{1-Z} \right), ~~ \s_2 = 2 \,\text{Im}\, \ln \left(1+ \sqrt{1-Z} \right) + \pi .
\eeq
\end{subequations}
Then, it follows that the SC mapping from the individual local coordinate patch with the local coordinate, $\zeta_r = \xi_r + i \eta_r$, $r=1, 2, 3, 4$ onto the complex plane can be written explicitly as 
\begin{subequations} 
\beq
e^{-\zeta_1} &=&  e^{\t_1+ i\s_1} \frac{(z-1)}{z(z-Z)},~~ e^{-\zeta_2} = -e^{\t_1+ i\s_1} \frac{(z-1)}{z(z-Z)},  \\
e^{-\zeta_3} &=&  e^{-\t_2 -i\s_2} \frac{z(z-Z)}{z-1},~~ e^{-\zeta_4} =  - e^{-\t_2-i\s_2} \frac{z(z-Z)}{z-1}.
\eeq
\end{subequations}

\section{Four-Closed String Scattering Amplitude}

By integrating $X^\m$ in Eq.(\ref{W4}), we can write the four-closed string scattering amplitude in 
terms of the momentum variables of four external strings, $P^{(r)}$, $r=1,2,3,4$ as follows 
\beq
{\cal W}_{[4]} &=& g^2  \langle  {\bf P}\vert 
\exp \,\Biggl\{\sum_{r=1}^4 \frac{\xi_r}{2} \left( \left(p^r_L\right)^2 + \left(p^r_R\right)^2 \right)+ 
\frac{1}{4} \sum_{r,s} \sum_{n, m } \bar C^{rs}_{nm} e^{|n| \xi_r + |m| \xi_s}\, P^{(r)}_n \cdot P^{(s)}_m \Biggr\} \vert 0 \rangle 
\eeq  
where $\vert  {\bf P}\rangle$ is the momentum eigenstate and $\vert 0 \rangle $ is the vacuum state in the oscillatory basis. 
Here, $\bar C^{rs}_{nm}$ is the Fourier component of the Green's function on the string worldsheet, analogous to the Neumann function
for open strings 
\beq
\ln \vert z_r - \zp_s \vert 
&=& - \d_{rs} \left\{\sum_{n=1} \frac{e^{-n\D}}{2n} \left(e^{in(\eta^\prime_s- \eta_r)} + e^{-in(\eta^\prime_s- \eta_r)} \right) - \max(\xi, \xi^\prime)  \right\} \nn\\
&& + \sum_{n, m} \bar C^{rs}_{nm} e^{|n| \xi_r + |m| \xi^\prime_s} e^{in\eta_r} e^{im\eta^\prime_s}
\eeq
where $\Delta = \vert \xi_r - \xi^\prime_s \vert$. 
The four-closed string vertex $ \vert V_{[4]} \rangle$ can be identified by 
\beq
{\cal W}_{[4]} &=& \langle  {\bf P} \vert \exp\left( \sum_{r=1}^4 \xi_r L^{(r)}_0 \right) \vert V_{[4]} \rangle . 
\eeq 
The Fock space representation of the four-closed string vertex is given by 
\beq
\vert V_{[4]} \rangle &=& e^{-\sum_{r=1}^4 2 \bar C^{rr}_{00}} 
\exp \Biggl\{\frac{1}{4} \sum_{r,s} \sum_{n, m \ge 1} \Biggl(\bar C^{rs}_{nm}\,\tilde\a^{(r)\dag}_n \cdot \tilde\a^{(s)\dag}_m +\bar C^{rs}_{-n-m} \, \a^{(r)\dag}_n \cdot \a^{(r)\dag}_m \Biggl) + \frac{1}{4} \sum_{r,s} \Biggr(\sum_{n \ge 1} \Bigr(
\bar C^{rs}_{n0}\tilde\a^{(r)\dag}_n\cdot p^{(s)}\nn\\
&& + \bar C^{rs}_{-n0} \a^{(r)\dag}_n\cdot p^{(s)} \Bigr)
+ \sum_{m \ge 1} p^{(r)} \cdot \Bigl(\bar C^{rs}_{0m}\tilde \a^{(s)\dag}_m +
\bar C^{rs}_{0-m} \a^{(s)\dag}_m \Bigr) + \bar C^{rs}_{00} p^{(r)} \cdot p^{(s)} 
\Biggr)\Biggr\} \vert 0 \rangle  .
\eeq 
The explicit forms of the Neumann functions for closed string, $\bar C^{rs}_{nm}$ are given by 
\begin{subequations}
\beq
\bar C^{rs}_{00} &=&  \ln \vert Z_r -Z_s \vert ,~~~ r\not=s, \\
\bar C^{rr}_{00} &=&  -\sum_{i\not=r} \frac{\a_i}{\a_r} \ln \vert Z_r - Z_i \vert + \frac{1}{\a_r} \t^{(r)}_0, \\
\bar C^{rs}_{n0} &=& \bar C^{rs}_{-n0} = \frac{1}{2n} \oint_{Z_r} \frac{d z}{2\pi i} \frac{1}{z-Z_s} e^{-n\zeta_r(z)}, ~~~n \ge 1,, \\
\bar C^{rs}_{nm} &=& \bar C^{rs}_{-n-m} =\frac{1}{2nm} \oint_{Z_r} \frac{dz}{2\pi i} \oint_{Z_s} \frac{d \zp}{2\pi i} 
\frac{1}{(z-\zp)^2} e^{-n\zeta_r(z) - m \zeta^\prime_s(\zp)}, ~~ n, m \ge 1 , \\
\bar C^{rs}_{n-m} &=& \bar C^{rs}_{-n m} = 0, ~~ n, m \ge 1  
\eeq 
\end{subequations}
By introducing the $SL(2,C)$ invariant measure for the Koba--Nielsen variables, the four-closed scattering amplitude 
can be written as follows:
\beq \label{4amp}
{\cal A}_{[4]} &=& g^2 \int \prod_{r=1}^4  dZ^2_r \frac{|Z_a-Z_b|^2|Z_b-Z_c|^2|Z_c-Z_a|^2}{d^2Z_a d^2Z_b d^2Z_c} \langle \Psi_{[4]} \vert V_{[4]} \rangle
\eeq 
where $\vert \Psi_{[4]} \rangle$ denotes the external four-closed string state.  

\section{Four-Graviton Scattering Amplitude}

The massless closed string state, $h_{\m\n} \a^\m_{1-} \tilde \a^\n_{-1} \vert 0 \rangle$ may be decomposed into graviton, anti-symmetric tensor, and scalar states. By choosing the symmetric traceless part of $h_{\m\n}$, the external four-graviton state can be written as 
\beq \label{4graviton}
\vert \Psi_{[4]} \rangle = \prod_{r=1}^4 \left\{h_{\m\n}(p^{(r)}) \a^{(r)\m}_{-1} \tilde \a^{(r)\n}_{-1} 
\right\}\vert 0 \rangle .
\eeq 
The graviton field $h^{\m\n}(p^{(r)})$ is subjected to the covariant gauge condition: $h^{\m\n}(p^{(r)}) p^{(r)}_\m = h^{\m\n}(p^{(r)}) p^{(r)}_\n = 0$. The four-graviton scattering amplitude follows from Eq. (\ref{4amp}) and Eq. (\ref{4graviton}):
\beq
{\cal A}_{[4G]} &=& \frac{g^2 c_{[4]}}{4}\int \vert Z_4 \vert^4 d^2 Z\, \prod_{r<s} \vert Z_r - Z_s \vert^{2 \,\frac{p^{(r)}}{2} \cdot \frac{p^{(s)}}{2}} e^{-2\sum_{r=1}^4 \bar C^{[4]rr}_{00}} 
\langle 0 \vert \left\{\prod_{i=1}^4 h_{\m\n}(p^{(i)}) a^{(i)\m}_1 \tilde a^{(i)\n}_1 \right\} \nn\\
&&
\Biggl\{ \left(\sum_{r,s} \bar C^{rs *}_{11} a^{(r)\dag}_1 \cdot a^{(s)\dag}_1
\right)^2  + \left(\sum_{r,s} \bar C^{rs *}_{11} a^{(r)\dag}_1 \cdot a^{(s)\dag}_1\right)\left(\sum_{r,s} \bar C^{rs*}_{10} a^{(r)\dag}_1 \cdot p^{(s)} \right)^2
\nn\\
&& + \frac{1}{4!}\left(\sum_{r,s} \bar C^{rs*}_{10} a^{(r)\dag}_1 \cdot p^{(s)} \right)^4
\Biggr\} \Biggl\{ \left(\sum_{r,s} \bar C^{rs }_{11} \,\tilde a^{(r)\dag}_1 \cdot \tilde a^{(s)\dag}_1\right)^2 \nn\\
&& + \left(\sum_{r,s} \bar C^{rs }_{11} \,\tilde a^{(r)\dag}_1 \cdot \tilde a^{(s)\dag}_1\right)\left(\sum_{r,s} \bar C^{rs}_{10} \tilde a^{(r)\dag}_1 \cdot p^{(s)} \right)^2
+ \frac{1}{4!}\left(\sum_{r,s} \bar C^{rs}_{10} \tilde a^{(r)\dag}_1 \cdot p^{(s)} \right)^4\Biggr\} \vert 0 \rangle 
\eeq 
where $c_{[4]}$ is a normalization constant, which will be fixed later. 
The following are explicit expressions of some $C$-functions that will be useful in evaluating ${\cal A}_{[4G]}$:
\begin{subequations} 
\beq
e^{-\bar C^{11}_{00}} &=& \frac{|Z|}{|Z_4|} e^{-\t_1}, ~~ e^{-\bar C^{22}_{00}} = \frac{|Z|}{|Z-1||Z_4|} e^{-\t_1},
~~e^{-\bar C^{33}_{00}} = \frac{|Z_4|}{|Z-1|} e^{\t_2}, ~~ e^{-\bar C^{44}_{00}} = \frac{1}{|Z_4|} e^{\t_2} , \\
\bar C^{11}_{10} &=& \frac{e^{\t_1 + i\s_1}}{2} \frac{1-Z}{Z^2}, ~~
\bar C^{12}_{10} = - \frac{e^{\t_1 + i\s_1}}{2} \frac{1}{Z^2}, ~~
\bar C^{13}_{10} = - \frac{e^{\t_1 + i\s_1}}{2} \frac{1}{Z}, ~~ 
\bar C^{14}_{10} = 0,\\
\bar C^{21}_{10} &=& \frac{e^{\t_1 + i\s_1}}{2} \frac{1-Z}{Z^2},~~ 
\bar C^{22}_{10} = -\frac{e^{\t_1 + i\s_1}}{2} \frac{1}{Z^2}, ~~
\bar C^{23}_{10} = -\frac{e^{\t_1 + i\s_1}}{2} \frac{1}{Z},~~
\bar C^{24}_{10} =0 , \\
\bar C^{31}_{10} &=& \frac{e^{-\t_2 - i\s_2}}{2} (1-Z),~~ 
\bar C^{32}_{10} = \frac{e^{-\t_2 - i\s_2}}{2}, ~~
\bar C^{33}_{10} = \frac{e^{-\t_2 - i\s_2}}{2} (2-Z),~~
\bar C^{34}_{10} =0, \\
\bar C^{41}_{10} &=& \frac{e^{-\t_2 - i\s_2}}{2} (1-Z),~~
\bar C^{42}_{10} = \frac{e^{-\t_2 - i\s_2}}{2},~~
\bar C^{43}_{10} = \frac{e^{-\t_2 - i\s_2}}{2} (2-Z),~~ 
\bar C^{44}_{10} =0, \\
\bar C^{12}_{11} &=& \frac{e^{2\t_1+ i2\s_1}}{2} \frac{(1-Z)}{Z^4}, ~~ 
\bar C^{13}_{11} = \frac{e^{\t_1+i\s_1 - \t_2 -i\s_2}}{2} \frac{(1-Z)}{Z}, ~~
\bar C^{14}_{11} = \frac{e^{\t_1+i\s_1 - \t_2 -i\s_2}}{2} \frac{1}{Z} ,\\
\bar C^{23}_{11} &=& \frac{e^{\t_1+i\s_1 - \t_2 -i\s_2}}{2} \frac{1}{Z}, ~~
\bar C^{24}_{11} = \frac{e^{\t_1+i\s_1 - \t_2 -i\s_2}}{2} \frac{(1-Z)}{Z}, ~~
\bar C^{34}_{11} = \frac{e^{-2\t_2-i2\s_2}}{2} (1-Z).
\eeq 
\end{subequations}

By ignoring some higher order terms in the expansion of $\ap$, we obtain the following expression for the four-graviton scattering amplitude:
\beq
{\cal A}_{[4G]} &=& g^2 c_{[4]} \int d^2 Z \,\vert Z\vert^{-\frac{s}{4}} \vert 1-Z\vert^{-\frac{t}{4}}
h_{\m_1\n_1} h_{\m_2\n_2} h_{\m_3\n_3} h_{\m_4\n_4}\nn\\
&& \Biggl\{ \frac{1}{Z^{*2}}\eta^{\m_1\m_2} \eta^{\m_3\m_4} + \eta^{\m_1\m_3} \eta^{\m_2\m_4}+  \frac{1}{(1-Z^*)^2}\, \eta^{\m_1\m_4} \eta^{\m_2\m_3} \nn\\
&& - \eta^{\m_1 \m_2} \frac{1}{4} \frac{1}{(1-Z^*)Z^{*2}} \left(Z^* p^{(1)\m_3} + p^{(4)\m_3}\right)
\left(Z^*p^{(2)\m_4} + p^{(3)\m_4}\right) \nn\\
&& + \eta^{\m_1 \m_3} \frac{1}{4} \frac{1}{(1-Z^*)Z^{*}} \left(p^{(1)\m_2} + Z^* p^{(4)\m_2}\right)
\left(Z^*p^{(2)\m_4} + p^{(3)\m_4}\right) \nn\\
&& - \eta^{\m_1 \m_4} \frac{1}{4} \frac{1}{(1-Z^*)^2 Z^*} \left((1-Z^*)p^{(4)\m_2} + p^{(3)\m_2}\right)
\left((1-Z^*) p^{(1)\m_3} + p^{(2)\m_3}\right) \nn\\
&& - \eta^{\m_2 \m_3} \frac{1}{4} \frac{1}{(1-Z^*)^2 Z^*} \left((1-Z^*) p^{(3)\m_1} + p^{(4)\m_1}\right)
\left(p^{(1)\m_4} + (1-Z^*) p^{(2)\m_4} \right) \nn\\
&& + \eta^{\m_2 \m_4} \frac{1}{4} \frac{1}{(1-Z^*) Z^*} \left(Z^* p^{(3)\m_1} + p^{(2)\m_1}\right)
\left(Z^* p^{(1)\m_3} + p^{(4)\m_3}\right) \nn\\
&& - \eta^{\m_3 \m_4} \frac{1}{4} \frac{1}{(1-Z^*)Z^{*2}} \left(Z^* p^{(3)\m_1} + p^{(2)\m_1}\right)
\left(p^{(1)\m_2} + Z^* p^{(4)\m_2}\right)\Biggr\} \nn\\
&&\Biggl\{ Z^* \Rightarrow Z, ~ \m \Rightarrow \n \Biggr\}
\eeq
where $s, t, u$ denote the Mandelstam variables defined as 
\beq
s = -\left(p^{(1)} + p^{(2)} \right)^2,~~t = -\left(p^{(2)} + p^{(3)} \right)^2,~~u = -\left(p^{(1)} + p^{(3)} \right)^2.
\eeq 
For four-graviton scattering, $s+t+u = 0$. If the integrand is expanded, it may contain as many as $27^2 = 729$ terms;
hence, it may be laborious to calculate the four-graviton scattering amplitude explicitly. 
However, owing to the useful formula found by Kawai, Lewellen, and Tye \cite{Kawai1986}, the integrand does not need to be expanded. Note that the terms in the integrand take the following form
\beq \label{integral}
I &=&\int d^2 Z \prod_{r<s} \vert Z_r - Z_s \vert^{2 \,\frac{p^{(r)}}{2} \cdot \frac{p^{(s)}}{2}} (Z^*)^{-n} 
(1-Z^*)^{-p}Z^{-m} (1-Z)^{-q} 
\eeq  
where $n, p, m, q$ are integers, and $I$ can be factorized into two integrals with two independent 
real variables, $\eta$ and $\xi$:
\begin{subequations}
\beq
I &=& \sin \left(\frac{\pi t}{8} \right) I_1 (n, p ) I_2 (m,q), \label{factor}\\
I_1(n,p) &=& \int_1^\infty d\eta \, \vert\eta\vert^{-\frac{s}{8}}
\vert 1- \eta \vert^{-\frac{t}{8}} \eta^{-n} \left(1-\eta \right)^{-p} \nn\\
&=&  (-1)^{p}\frac{\Gamma\left(-\frac{u}{8}+n +p -1\right) \Gamma\left(-\frac{t}{8} -p +1 \right)}
{\Gamma\left(\frac{s}{8} +n\right)}, \label{factor1} \\
I_2(m,q) &=& 
\int_0^1 d \xi\, \vert \xi\vert^{-\frac{s}{8}} \vert 1- \xi \vert^{-\frac{t}{8}} \xi^{-m} \left(1- \xi\right)^{-q} \nn\\
&=& \frac{\Gamma\left(-\frac{s}{8}-m +1\right) \Gamma\left(-\frac{t}{8} -q +1 \right)}
{\Gamma\left(\frac{u}{8} -m-q+2\right)} \label{factor2} . 
\eeq 
\end{subequations}
From Eq. (\ref{integral}) and Eqs. (\ref{factor},\ref{factor1},\ref{factor2}), we find  
\beq \label{obtained}
{\cal A}_{[4G]\text{proper}} &=& g^2 c_{[4]} h_{\m_1\n_1} h_{\m_2\n_2} h_{\m_3\n_3} h_{\m_4\n_4} \sin \left(\frac{\pi t}{8}\right) \nn\\
&& \Biggl\{I_1(2,0) \eta^{\m_1 \m_2} \eta^{\m_3 \m_4} + I_1(0,0) \eta^{\m_1 \m_3} \eta^{\m_2 \m_4} +
I_1(0,2) \eta^{\m_1 \m_4} \eta^{\m_2 \m_3} \nn\\
&&- \frac{1}{4}\eta^{\m_1 \m_2} \left(I_1(0, 1) p^{(1)\m_3} p^{(2)\m_4}+ I_1(1, 1)p^{(1)\m_3} p^{(3)\m_4}+ I_1(1, 1) p^{(4)\m_3} p^{(2)\m_4}+ I_1(2, 1) p^{(4)\m_3} p^{(3)\m_4}\right)  \nn\\
&&+ \frac{1}{4}\eta^{\m_1 \m_3} \left(I_1(0, 1) p^{(1)\m_2} p^{(2)\m_4}+ I_1(1, 1) p^{(1)\m_2} p^{(3)\m_4}+ I_1(-1, 1) p^{(4)\m_2} p^{(2)\m_4}+ I_1(0, 1) p^{(4)\m_2} p^{(3)\m_4}\right)\nn\\
&&- \frac{1}{4}\eta^{\m_1 \m_4} \left(I_1(1, 0) p^{(4)\m_2} p^{(1)\m_3}+ I_1(1, 1) p^{(4)\m_2} p^{(2)\m_3}+ I_1(1, 1) p^{(3)\m_2} p^{(1)\m_3}+ I_1(1, 2) p^{(3)\m_2} p^{(2)\m_3}\right)\nn\\
&&- \frac{1}{4}\eta^{\m_2 \m_3} \left(I_1(1, 0) p^{(3)\m_1} p^{(2)\m_4}+ I_1(1, 1) p^{(3)\m_1} p^{(1)\m_4}+ I_1(1, 1) p^{(4)\m_1} p^{(2)\m_4}+ I_1(1, 2) p^{(4)\m_1} p^{(1)\m_4}\right)\nn\\
&&+ \frac{1}{4}\eta^{\m_2 \m_4} \left(I_1(-1, 1) p^{(3)\m_1} p^{(1)\m_3}+ I_1(0,1) p^{(3)\m_1} p^{(4)\m_3}+ I_1(0, 1) p^{(2)\m_1} p^{(1)\m_3}+ I_1(1, 1) p^{(2)\m_1} p^{(4)\m_3}\right)\nn\\
&&- \frac{1}{4}\eta^{\m_3 \m_4} \left(I_1(0, 1) p^{(3)\m_1} p^{(4)\m_2}+ I_1(1, 1) p^{(3)\m_1} p^{(1)\m_2}+ I_1(1, 1) p^{(2)\m_1} p^{(4)\m_2}+ I_1(2, 1) p^{(2)\m_1} p^{(1)\m_2}\right)
\Biggr\} \nn\\
&&\Biggl\{ I_1(n,p) \Rightarrow I_2(n,p),  ~ \m \Rightarrow \n \Biggr\}
\eeq
To compare the obtained graviton scattering amplitude Eq. (\ref{obtained}) with that of the conventional string theory, we can rewrite it as follows: 
\beq \label{Aproper}
{\cal A}_{[4G]\text{proper}} &=& \frac{\kappa^2}{128} 
\frac{\G\left(-\frac{s}{8}\right)\G\left(-\frac{t}{8}\right)\G\left(-\frac{u}{8}\right)}
{\G\left(1+ \frac{s}{8}\right)\G\left(1+ \frac{t}{8}\right)\G\left(1+ \frac{u}{8}\right)}
h_{\m_1\n_1} h_{\m_2\n_2} h_{\m_3\n_3} h_{\m_4\n_4} 
K^{\m_1\m_2\m_3\m_4}_{\rm proper} K^{\n_1\n_2\n_3\n_4}_{\rm proper}.
\eeq
Here, $\kappa^2 = - c_{[4]} \pi g^2/2$, and $K^{\m_1\m_2\m_3\m_4}_{\text{proper}}$ is 
the kinematic factor for four-gauge-particle scattering amplitude of 
open string theory in the proper-time gauge in Ref. \cite{TLeeGauge2018},
\beq \label{proper}
K^{\m_1\m_2\m_3\m_4}_{\text{proper}} &=& -\frac{ut}{4\left(1+ \frac{s}{8}\right)} \eta^{\m_1 \m_2} \eta^{\m_3 \m_4} 
- \frac{st}{4\left(1+ \frac{u}{8}\right)} \eta^{\m_1 \m_3} \eta^{\m_2 \m_4} 
- \frac{us}{4\left(1+ \frac{t}{8}\right)}  \eta^{\m_1 \m_4} \eta^{\m_2 \m_3} \nn\\
&&+ \frac{1}{2}\eta^{\m_1 \m_2} \Biggl( - s p^{(1)\m_3} p^{(2)\m_4}+ u p^{(1)\m_3} p^{(3)\m_4}+ 
u p^{(4)\m_3} p^{(2)\m_4}+ \frac{u
\left(1-\frac{u}{8}\right)}{\left(1+\frac{s}{8}\right)} p^{(4)\m_3} p^{(3)\m_4}\Biggr)  \nn\\
&&- \frac{1}{2}\eta^{\m_1 \m_3} \left(- s p^{(1)\m_2} p^{(2)\m_4}
+ u p^{(1)\m_2} p^{(3)\m_4}- s p^{(4)\m_2} p^{(3)\m_4}
- \frac{ s\left(1-\frac{s}{8}\right)}{\left(1+\frac{u}{8}\right)}p^{(4)\m_2} p^{(2)\m_4}
\right)\nn\\
&&+ \frac{1}{2}\eta^{\m_1 \m_4} \left( - t p^{(4)\m_2} p^{(1)\m_3}+ u p^{(4)\m_2} p^{(2)\m_3}
+ u p^{(3)\m_2} p^{(1)\m_3}+ \frac{u 
\left(1-\frac{u}{8}\right)}{\left(1+\frac{t}{8}\right)} p^{(3)\m_2} p^{(2)\m_3}\right)\nn\\
&&+ \frac{1}{2}\eta^{\m_2 \m_3} \left(- t
p^{(3)\m_1} p^{(2)\m_4}+ u p^{(3)\m_1} p^{(1)\m_4}+ u p^{(4)\m_1} p^{(2)\m_4}+ 
\frac{u
\left(1-\frac{u}{8}\right)}{\left(1+\frac{t}{8}\right)} p^{(4)\m_1} p^{(1)\m_4}\right)\nn\\
&&- \frac{1}{2}\eta^{\m_2 \m_4} \left( - s p^{(3)\m_1} p^{(4)\m_3}- s p^{(2)\m_1} p^{(1)\m_3}
+ u p^{(2)\m_1} p^{(4)\m_3}-\frac{ s
\left(1-\frac{s}{8}\right)}{\left(1+\frac{u}{8}\right)}
p^{(3)\m_1} p^{(1)\m_3}\right)\nn\\
&&+ \frac{1}{2}\eta^{\m_3 \m_4} \left(- s p^{(3)\m_1} p^{(4)\m_2}+ u p^{(3)\m_1} p^{(1)\m_2}+ 
u p^{(2)\m_1} p^{(4)\m_2}+ \frac{u
\left(1-\frac{u}{8}\right)}{\left(1+\frac{s}{8}\right)} p^{(2)\m_1} p^{(1)\m_2}\right).
\eeq 
Note that because the kinematic factor in the proper-time gauge, $K^{\m_1\m_2\m_3\m_4}_{\text{proper}}$
has tachyon poles in all three channels, the four-graviton-scattering amplitude in the proper-time gauge ${\cal A}_{[4G]\text{proper}}$ Eq. (\ref{Aproper}) has simple poles on the Regge trajectories,
\beq
\frac{s}{8}, \frac{t}{8}, \frac{u}{8} = -1, 0, 1, 2, 3 , \dots , 
\eeq 
in a manner consistent with the cubic couplings between tachyon and graviton fields. It follows 
from the cubic coupling of one tachyon and two graviton fields, $\phi h_{\m\n} h^{\m\n}$ that 
the four-graviton-scattering amplitude must have tachyon poles.    

The four-graviton-scattering amplitude obtained by the conventional vertex operator technique
\cite{Schwarz1982} 
is also written in the same form with different kinematic factor:
\begin{subequations}
\beq \label{vertex}
{\cal A}_{[4G]\text{vertex}} &=& \frac{\kappa^2}{128} 
\frac{\G\left(-\frac{s}{8}\right)\G\left(-\frac{t}{8}\right)\G\left(-\frac{u}{8}\right)}
{\G\left(1+ \frac{s}{8}\right)\G\left(1+ \frac{t}{8}\right)\G\left(1+ \frac{u}{8}\right)}
h_{\m_1\n_1} h_{\m_2\n_2} h_{\m_3\n_3} h_{\m_4\n_4} 
K^{\m_1\m_2\m_3\m_4} K^{\n_1\n_2\n_3\n_4}, \\
K^{\m_1\m_2\m_3\m_4} &=&  -\frac{tu}{4}\eta^{\m_1 \m_2} \eta^{\m_3 \m_4} - \frac{st}{4} \eta^{\m_1 \m_3} \eta^{\m_2 \m_4} -  \frac{us}{4} \eta^{\m_1 \m_4} \eta^{\m_2 \m_3} \nn\\
&& + \frac{\eta^{\m_1 \m_2}}{2} \left(t p^{(1)\m_3} p^{(2)\m_4}+ u p^{(2)\m_3} p^{(1)\m_4} \right)  \nn\\
&& + \frac{\eta^{\m_1 \m_3}}{2} \left(t p^{(1)\m_2} p^{(3)\m_4}+ s p^{(3)\m_2} p^{(1)\m_4} \right)  \nn\\
&& + \frac{\eta^{\m_1 \m_4}}{2} \left(u p^{(1)\m_2} p^{(4)\m_3}+ s p^{(4)\m_2} p^{(1)\m_3} \right)  \nn\\
&& + \frac{\eta^{\m_2 \m_3}}{2} \left(u p^{(2)\m_1} p^{(3)\m_4}+ s p^{(3)\m_1} p^{(2)\m_4}  \right)  \nn\\
&& + \frac{\eta^{\m_2 \m_4}}{2} \left(t p^{(2)\m_1} p^{(4)\m_3}+s p^{(4)\m_1} p^{(2)\m_3}  \right)  \nn\\
&& + \frac{\eta^{\m_3 \m_4}}{2} \left(t p^{(3)\m_1} p^{(4)\m_2}+ u p^{(4)\m_1} p^{(3)\m_2}  \right).
\eeq
\end{subequations}
Both four-graviton-scattering amplitudes, Eq. (\ref{Aproper}) and Eq. (\ref{vertex})  
reduce to that of Einstein’s gravity in the zero-slope limit \cite{Sannan1986},
which only has massless poles: 
\beq
{\cal A}_{[4G]{\rm Einstein}} &=& 4 \kappa^2 h_{\m_1\n_1} h_{\m_2\n_2} h_{\m_3\n_3} h_{\m_4\n_4} \frac{1}{stu}
K^{\m_1\m_2\m_3\m_4}K^{\n_1\n_2\n_3\n_4}.
\eeq 
However, in contrast to the four-graviton-scattering amplitude in the proper-time gauge Eq. (\ref{Aproper}), the four-graviton-scattering amplitude, Eq. (\ref{vertex}) does not contain tachyon poles 
.
The scattering amplitude ${\cal A}_{[4G]\text{vertex}}$ has simple poles where 
\beq
\frac{s}{8}, \frac{t}{8}, \frac{u}{8} = 0, 1, 2, 3 , \dots .
\eeq 
Therefore, at finite $\ap$, both graviton scattering amplitudes differ. 
Because the four-graviton scattering amplitude ${\cal A}_{[4G]{\rm proper}}$ 
Eq. (\ref{Aproper}) is obtained by evaluating the Polyakov string path integral without any approximation,
it is valid for the entire energy scale. Therefore, it may be more appropriate to 
use the scattering amplitude ${\cal A}_{[4G]{\rm proper}}$ than others to investigate the short distance behaviors of quantum gravity.

\section{Conclusions}

In the proper-time gauge, the Polyakov string path integral can be expressed in a form similar to the Schwinger's proper-time representation of the Feynman integral of the quantum field theory so that the second quantized expressions of string scattering amplitudes can be identified. 
We have shown that the four-graviton scattering amplitude of Einstein’s gravity arises as a zero-slope limit of the four-closed string scattering amplitude by explicitly evaluating the corresponding string path integral in the covariant proper-time gauge. 
According to the general studies on gravitons \cite{Feynman1963,WeinbergPL1964,WeinbergPR1964,WeinbergPR1965,Deser1970,Boulware1975}, all consistent theories of graviton must reduce to Einstein’s gravity in the low energy limit. We confirmed that this general argument applies to closed string theory, containing the massless spin-two particle state.

However, if string corrections become
relevant with finite $\ap$, the four-graviton-scattering amplitude differs from that of Einstein’s 
gravity significantly. Moreover, it considerably differs from the conventional graviton
scattering amplitude of the string theory, which approximates the external strings using local vertex operators. A notable point is that the former obtained in the present work 
contains all the Regge poles of the closed string, 
including the tachyon poles in a manner consistent with the cubic 
couplings between tachyon and graviton fields. However, the latter, adopting the vertex operator 
technique, is missing the tachyon poles. These tachyon poles may be important in resolving 
important problems associated with the short distance behavior of gravity such as the dark matter 
problem \cite{Bertone2018} or UV divergence of perturbative quantum gravity \cite{Goroff1976}, 
because it may considerably alter the high energy behavior of quantum gravity. 

In Ref. \cite{TLeePLB2017}, three-closed-string scattering amplitudes were shown to be completely factorized into those of three-open-string amplitudes, and the Neumann functions for closed string were found to be the same as those of the open string for three-closed-string scattering. This may imply that the KLT relations in the first quantized theory may be fully extended to the second quantized string theory. Here, we further extend the work on the three-graviton-scattering amplitudes to the four-graviton-scattering amplitudes, showing that the four-graviton scattering amplitude of 
A closed string can be completely factorized. By showing that factorization works even at finite $\ap$, we can say that $\ap$-the corrections in closed string theory can be entirely determined by $\ap$-corrections in the open string theory. This work may be important for generalizing the Bern-Carrasco--Johansson duality \cite{BernPRL2010}, which relates the scattering amplitudes of non-Abelian gauge particles with those of gravitons within the framework of the second quantized closed string theory in the proper-time gauge.

\vskip 1cm

\begin{acknowledgments}

This work was supported by Basic Science Research Program through the National Research Foundation of Korea (NRF) funded by the Ministry of Education (2017R1D1A1A02017805).
The author thanks J. C. Lee and Y. Yang for useful discussions and their hospitality during author's visit to NCTU (Taiwan) and also acknowledges the the hospitality at APCTP where 
part of this work was done. 
\end{acknowledgments}


%

%





\end{document}